\documentclass[12pt]{JHEP3}
\usepackage{epsfig,amsmath,amssymb,graphics}
\newcommand{\hc}{\text{ h.c}}
\newcommand{\Tr}{\text{\,Tr\,}}
\newcommand{\TeV}{\text{ TeV }}

\title{Precision Electroweak Observables in the Minimal Moose Little Higgs Model}

\author{Can Kilic and Rakhi Mahbubani\\
Jefferson Physical Laboratory, Harvard University, Cambridge, MA 02138\\
E-mail: \email{kilic@fas.harvard.edu}, \email{rakhi@physics.harvard.edu}}

\preprint{hep-ph/0312053\\HUTP-03/A111}

\abstract{Little Higgs theories, in which the Higgs particle is realized as the
pseudo-Goldstone boson of an approximate global chiral symmetry
have generated much interest as possible alternatives to weak
scale supersymmetry. In this paper we analyze precision electroweak observables in the Minimal Moose model and find that in order to be consistent with current experimental bounds, the gauge structure of this theory needs to be modified.  We then look for viable regions of
parameter space in the modified theory by calculating the various contributions to the S and T parameters.}

\begin{document}

\section{Introduction}
Despite its spectacular agreement with current experimental
data, the Standard Model (SM) is widely held to be incomplete due
to an instability in its Higgs sector; radiative corrections to
the Higgs mass suffer from one-loop quadratic divergences leading to 
an undesirable level of fine-tuning between
the bare mass and quantum corrections. This suggests the emergence
of new physics at energy scales around a TeV, which will be
investigated in the near future with direct accelerator searches. The 
electroweak sector of the SM has been probed to better than the 1\% level 
by precision experiments at low energies as well as at the $Z$-pole by LEP 
and SLC.  The data obtained can also severely constrain possible extensions of the 
SM at TeV energies \cite{Hagiwara:2002fs,Burgess:1994vc,Altarelli:1998et,Hagiwara:1994pw}.

Recently, a new class of theories known as Little Higgs (LH) models
\cite{Arkani-Hamed:2002qx,Arkani-Hamed:2002qy,Gregoire:2003kr,Chang:2003zn,Kaplan:2003uc,Low:2002ws,Skiba:2003yf}
have been proposed to understand the lightness of the Higgs by making it a 
pseudo-Goldstone boson.
Approximate global symmetries ensure the cancellation of all
quadratic sensitivity to the cutoff at one loop in the gauge, Yukawa and Higgs
sectors, by partners of the same quantum statistics: heavy gauge
bosons cancel the divergence of the SM gauge loop; massive scalars
do the same for the Higgs self-coupling, as do heavy fermions for
the top loop contributions. These partner particles have masses of
the order of the symetry breaking scale $f$, which we take to be a few TeV. At lower energies the presence of these new particles
can be felt only through virtual exchanges and and their effects on 
precision electroweak oberservables (PEWOs).  

In this paper we calculate corrections to PEWOs in the Minimal Moose 
(MM) \cite{Arkani-Hamed:2002qx}, and in a similar model with a slight 
variation in gauge structure (the Modified Minimal Moose, or MMM) in an attempt to find regions of parameter space where these are small with a tolerable level of fine tuning in the Higgs sector.

Both models have a simple product gauge structure
$G_L\times G_R$ and reduce to the SM with additional Higgs doublets at
low energies.  Above the symmetry breaking scale the Higgs sector is a nonlinear sigma model which becomes strongly coupled at
$\Lambda\simeq 10$ TeV and requires UV completion at higher energies.  
The enhanced gauge sector can contribute to precision observables through 
the interaction of the partners to SM gauge bosons,
$W'$ and $B'$, with fermions and Higgs doublets via currents
$j^{\mu}_{F}$ and $j^{\mu}_{H}$ respectively, generating low energy operators
of the form $j_{F}j_{F}$, $j_{F}j_{H}$ and $j_{H}j_{H}$.  We group these into 
oblique and non-oblique corrections, where the former impact precision experiments 
only via their effects on gauge boson propagators, and summarize their salient
properties below.

Oblique corrections can originate from:
\begin{itemize}
\item{Interactions $j_{H}j_{H}$\\
$B'$ exchange modifies the $Z_{0}$ mass and
hence introduces custodial $SU(2)$ violating effects to which the
$T$ parameter is sensitive. This is a cause for concern in the MM,
but is reduced considerably in the Modified Moose by gauging 
a different subset of the global symmetries.}

\item{Non-linear sigma model (nlsm) kinetic terms\\
At energies above the global symmetry breaking scale, the Higgs
doublets form components of nlsm fields with
self-interactions. This gives rise to custodial $SU(2)$ violating
operators in the low energy theory which become our most
significant constraint.}

\item{Higgs-heavy scalar interactions\\
The theories also contain a scalar potential in the form of
plaquette terms to ensure that electroweak symmetry is broken
appropriately. This contributes to the $T$ parameter through the
exchange of heavy scalar modes, which effect we show to be
negligible.}

\item{Fermion loops\\
The presence of a vector-like partner to the top quark is another
source for $T$  and $S$ parameter contributions. We calculate these and show
that they are tolerably small for a wide range of parameters of
the theory.}

\item{Higgs loops\\
Since the MM is a two Higgs doublet theory at low energies, corrections 
due Higgs loops are similar to those of the Minimal Supersymmetric 
Standard Model.}

\end{itemize}

The following are the non-oblique corrections of concern to us:
\begin{itemize}
\item{Four-fermion operators $j_{F}j_{F}$\\
These modify $G_{F}$ and can be controlled in the MMM in the
near-oblique limit (see below), in which light
SM fermions decouple from the $W'$ and $B'$.}

\item{Interactions $j_{F}j_{H}$\\
Operators of this form shift the coupling of the SM gauge bosons
to the fermions (most easily seen in unitary gauge). These are
also minimized in the near-oblique limit.}
\end{itemize}

LH gauge sectors generically have a simple limit in
which highly constrained non-oblique corrections vanish (the near-oblique limit  
\cite{Gregoire:2003kr})
in tandem with oblique corrections from the gauge boson sector. In
the MM however, the $SU(3)\times SU(2)\times U(1)$ gauge structure
is too tightly constrained to allow for a decrease in the large
oblique $B'$ correction by variation of the gauge couplings. This
issue is resolved in the MMM by replacing the $SU(3)$ gauge group
by another $SU(2)\times U(1)$ and charging the light fermions
equally under both $U(1)$s, giving
\begin{equation}\label{structure of gauge boson charges}
j_{H},j_{F_{light}}\propto\tan{\theta'}-\cot{\theta'}\nonumber
\end{equation}
for $\tan{\theta'}=g_{1R}/g_{1L}$, the ratio of the U(1) couplings at the
sites.  Setting these nearly equal to each other rids us of large 
heavy gauge boson contributions to the $T$ parameter as well 
as undesirable light four-fermion operators arising from $B'$ 
exchange. This method does not work with third-generation fermions 
which are coupled to the Higgs in a slightly different way.  Possible 
non-oblique corrections involving these will not be discussed 
since they are not yet unambiguously constrained by experiment.  
For additional discussion of this see \cite{Gregoire:2003kr}.  $W'$-
exchange operators are more easily handled since, provided we 
stay away from the strong coupling regime, increasing one of the 
SU(2) gauge couplings with respect to the other increases the 
mass of the $W'$, effectively decoupling it from our theory.

We begin this paper with a brief review of the MM, keeping as far
as possible to the conventions used in \cite{Arkani-Hamed:2002qx}.  
In Sections \ref{Sec: gauge bosons} to \ref{Sec: plaquettes} we 
calculate tree-level corrections to PEWOs from different sectors. We
go on to discuss electroweak symmetry breaking in the low energy
theory (Section \ref{Sec: EWSB}) and determine loop effects due to a new 
heavy fermion (Section \ref{Sec: fermion loops}) and Higgs doublets (Section \ref{Sec: 
higgs loops}). In spite of the fact that the MM is inconsistent with current constraints on PEWOs we show in Section \ref{Sec: results} that there are regions in the parameter space of the MMM where all except third
generation non-oblique corrections can be eliminated, with
tolerably small oblique corrections.  We will see that the most
unforgiving aspect of both models is the non-linear sigma model
sector which has no residual $SU(2)_c$ symmetry and hence
gives rise to a $T$ parameter contribution that can only be decreased
by adjusting $f$.  This compels us to choose $f\gtrsim
2$ TeV. We display two sets of parameters, one that is well within the 1.5-$\sigma$ $S$-$T$ ellipse with a 17\%  fine tuning in the SM Higgs mass and another that falls just outside the ellipse with a 3\% fine tuning.  We show that there are regions of parameter space where one can do even better than the first set, however this is only possible for a rather specific choice of parameters.  We measure
fine tuning by $(m/\delta m)^2$, where $\delta m$ is the top loop 
correction to the mass of the Higgs doublet, and $m$ is the physical 
Higgs mass.

\section{The Theory}

\FIGURE[ht]{ \epsfig{file=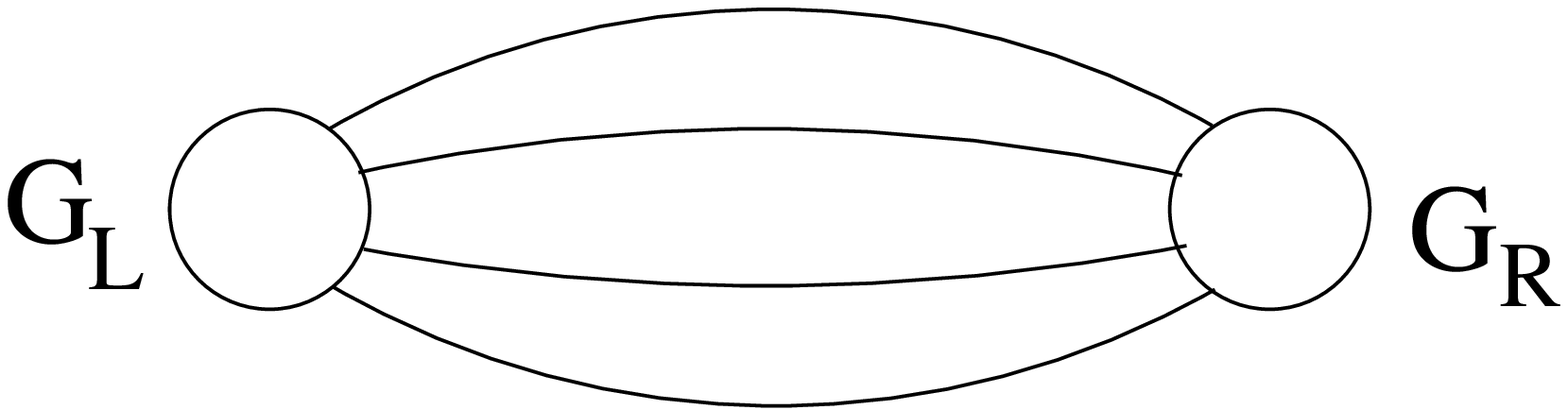, width = 7cm}
\caption{The Minimal Moose} \label{Fig: Minimal Moose}}

The Minimal Moose is a two-site four-link model with gauge
symmetry $G_{R}=SU(3)$ at one site and $G_{L}=SU(2)\times U(1)$ at
the other.  The standard model fermions are charged under $G_{L}$,
with their usual quantum numbers while the link fields
$X_{j}=\exp(2ix_{j}/f)$ are 3 by 3 nonlinear sigma model fields
transforming as bifundamentals under $G_{L}\times G_{R}$ where a
fundamental of $G_{L}$ is
$\textbf{2}_{1/6}\oplus\textbf{1}_{-1/3}$. These fields get
strongly coupled at a scale $\Lambda=4\pi f$, beneath which the
theory is described by the Lagrangian
\begin{equation}\label{equ: Lagrangian}
\mathcal{L}=\mathcal{L}_{G}+\mathcal{L}_{\chi}+\mathcal{L}_{t}+\mathcal{L}_{\psi}
\end{equation}
$\mathcal{L}_{G}$ includes all kinetic terms and gauge
interactions, while $\mathcal{L}_{\chi}$ contains plaquette
couplings between the $X_{j}$:
\begin{eqnarray}\label{equ: Plaquette terms}
\mathcal{L}_{\chi}=\left(\frac{f}{2}\right)^{4}\bigg(\Tr\left[A_{1}X_{1}X_{2}^{\dag}X_{3}X_{4}^{\dag}\right]+\Tr\left[A_{2}X_{2}X_{3}^{\dag}X_{4}X_{1}^{\dag}\right]\nonumber\\+\Tr\left[A_{3}X_{3}X_{4}^{\dag}X_{1}X_{2}^{\dag}\right]
+\Tr\left[A_{4}X_{4}X_{1}^{\dag}X_{2}X_{3}^{\dag}\right]\bigg)+\mathrm{h.c.}
\end{eqnarray}
with $A_{i} = \kappa_{i}+\epsilon_{i}T^{8}$ for $\epsilon\sim
\kappa/10$.  This is a natural relation since any radiative corrections
to $\epsilon$ require spurions from both the gauge and plaquette 
sectors and so can only arise at two loops.  The $\epsilon$ terms 
give the little Higgses a mass (see Equation \ref{equ: tree level Higgs masses})
and are required to stabilize electroweak symmetry breaking (EWSB).

The third generation quark doublet is coupled to a pair of colored Weyl
fermions $U$,$U^{c}$ via Yukawa terms in $\mathcal{L}_{t}$
\begin{equation}\label{equ: Yukawa couplings}
\mathcal{L}_{t}=\lambda
f\left(\begin{array}{ccc}0&0&u_{3}^{c'}\end{array}\right)X_{1}X_{4}^{\dag}\left(\begin{array}{c}q_{3}\\U\end{array}\right)+\lambda^{'}fUU^{c}+\hc
\end{equation}
and $\mathcal{L}_{\psi}$ contains the remaining Yukawa couplings.  These
take the same form as above for the light up-type quarks, but with $U$ and $U^{c}$
removed, while the down and charged lepton sectors look like
\begin{equation}\label{equ: light Yukawa couplings}
\mathcal{L}_{\psi}\supset\lambda_{D}\left(\begin{array}{cc}q&0\end{array}\right)X_{1}X_{4}^{\dag}\left(\begin{array}{c}0\\0\\d^{c}\end{array}\right)+\lambda_{E}\left(\begin{array}{cc}l&0\end{array}\right)X_{1}X_{4}^{\dag}\left(\begin{array}{c}0\\0\\e^{c}\end{array}\right)+\hc
\end{equation}

We also impose a $Z_{4}$ symmetry which cyclically permutes 
the link fields and hence requires equality of all the decay 
constants ($f_{i}=f$) and plaquette couplings ($\kappa_{i}=\kappa$,
$\epsilon_{i}=\epsilon$).  The only $Z_{4}$-breaking terms arise
in the fermion sector and are small.

\section{The Gauge Boson Sector}
\label{Sec: gauge bosons}

The link fields Higgs the $G_{L}\times G_{R}$ gauge groups down to
the diagonal $SU(2)\times U(1)$ subgroup, leaving one set each of
massive and massless gauge bosons. This can be seen explicitly by
considering the link field covariant derivatives:
\begin{equation}\label{equ: covariant derivatives for link fields}
\widetilde{D}_{\mu}X_{j}=\partial_{\mu}X_{j}-ig_{3}X_{j}A_{\mathbf{3,R}\mu}^{A}\mathbf{T}^{A}+ig_{2}A_{\mathbf{2,L}\mu}^{a}\mathbf{T}^{a}X_{j}+iqg_{1}A_{\mathbf{1,L}\mu}^{}\mathbf{T}^{8}X_{j}
\end{equation}
where the $\mathbf{T}$s for $A=1,...,8$ and $a=1,2,3$ are $SU(3)$ generators
normalized such that
$\Tr[\mathbf{T}^{A}\mathbf{T}^{B}]=\frac{\delta^{AB}}{2}$ (similarly for a,b indices); and
$q=1/\sqrt{3}$ to ensure that the Higgs doublet eventually has the
correct SM hypercharge. Expanding out the fields ($X_{j}=\exp(2ix_{j}/f)$) in the
kinetic term
\begin{equation}\label{equ: link field kinetic terms}
\frac{f^{2}}{4}\Tr\left[\sum_{j=1}^{4}(\widetilde{D}_{\mu}X_{j})(\widetilde{D}^{\mu}X_{j})^{\dagger}\right]
\end{equation}
shows that the eaten Goldstone boson, $w$, is proportional to
$x_{1}+x_{2}+x_{3}+x_{4}$.  Orthogonal combinations
$x$,$y$ and $z$ can be defined as follows:
\begin{equation}\label{equ: transformation matrix}
\left( \begin{array}{c} w\\ z\\ x\\ y
\end{array}\right)= \frac{1}{2} \left( \begin{array}{cccc} +1 & +1 & +1 &
+1\\ +1 & -1 & +1 & -1\\ -1 & -1 & +1 & +1\\ -1 & +1 & +1 & -1
\end{array}\right)\\ \left( \begin{array}{c} x_{1}\\ x_{2}\\ x_{3}\\ x_{4}
\end{array}\right)\\
\end{equation}
where each of the above fields decomposes under $SU(2)\times U(1)$ 
as $\textbf{3}_{0}$ $(\phi)$ + $\textbf{1}_{0}$ $(\eta) + \textbf{2}_{\pm1/2}$  $(h\text{ and }h^{\dag}).$
\begin{equation}\label{equ: 3 X 3 representation of PGB}
x=\left(\begin{array}{cc}\widetilde{\phi}_{x}+\frac{\eta_{x}}{\sqrt{12}}&\frac{h_{x}}{\sqrt{2}}\\\frac{h_{x}^{\dag}}{\sqrt{2}}&\frac{-\eta_{x}}{\sqrt{3}}\end{array}\right)
\end{equation}
The $x$ and $y$ contain two Higgs doublets in the more familiar form
\begin{eqnarray}\label{equ: transformation to $h_{1}h_{2}$ basis}
h_{1}=\frac{h_{x}+ih_{y}}{\sqrt{2}}\\
h_{2}=\frac{h_{x}-ih_{y}}{\sqrt{2}}
\end{eqnarray}
and plaquette terms give $z$ a large tree level mass, so it can
be integrated out of the theory at a TeV.

Going to unitary gauge results in a mass
matrix for heavy gauge bosons $W'_{\mu}$, $B'_{\mu}$ and
$A_{\mathbf{3}\mu}$ with eigenvalues $2gf/\sin{2\theta}$,
$2qg'f/\sin{2\theta'}$ and $fg/\sin{\theta}$ respectively.  The
$W'$ s and $B'$ s are admixtures of $A_{\mathbf{3,R}}$,
$A_{\mathbf{2,L}}$ and $A_{\mathbf{1,L}}$:
\begin{eqnarray}\label{equ: gauge boson mass eigenstates}
W_{\mu}^{a}&=&\cos{\theta}
A_{\mathbf{2,L}\mu}^{a}+\sin{\theta} A_{\mathbf{3,R}\mu}^{a}\nonumber\\
W'^{a}_{\mu}&=&-\sin{\theta}
A_{\mathbf{2,L}\mu}^{a}+\cos{\theta} A_{\mathbf{3,R}\mu}^{a}\nonumber\\
B_{\mu}&=&\cos{\theta'}
A_{\mathbf{1,L}\mu}+\sin{\theta'} A_{\mathbf{3,R}\mu}^{8}\\
B'_{\mu}&=&-\sin{\theta'} A_{\mathbf{1,L}\mu}+\cos{\theta'}
A_{\mathbf{3,R}\mu}^{8}\nonumber
\end{eqnarray}
with mixing angles defined as follows:
\begin{eqnarray}\label{equ: definitions of mixing angles}
g&=&\frac{g_{2}g_{3}}{\sqrt{g_{2}^{2}+g_{3}^{2}}}\nonumber\quad\quad\quad\quad\sin{\theta}=\frac{g}{g_{3}} \\
g'&=&\frac{g_{1}g_{3}}{\sqrt{\left(qg_{1}\right)^{2}+g_{3}^{2}}}\quad\quad\sin{\theta'}=\frac{qg'}{g_{3}}\nonumber
\end{eqnarray}
The Higgses couple to heavy gauge bosons via the following
currents:
\begin{eqnarray}\label{equ: gauge boson currents}
j^{a}_{W^{'}\mu}&=&-\frac{ig}{2\tan{2\theta}}\left[h_{1}^{\dag}\sigma^{a}\overleftrightarrow{D_{\mu}}h_{1}+h_{2}^{\dag}\sigma^{a}\overleftrightarrow{D_{\mu}}h_{2}\right]\nonumber\\
j_{B^{'}\mu}&=&-\frac{\sqrt{3}iqg^{'}}{2\tan{2\theta^{'}}}\left[h_{1}^{\dag}\overleftrightarrow{D_{\mu}}h_{1}+h_{2}^{\dag}\overleftrightarrow{D_{\mu}}h_{2}\right]
\end{eqnarray}
where $D_{\mu}$ is a Standard Model covariant derivative 
and $\sigma$s are Pauli matrices.

Explicitly integrating out the heavy gauge bosons results in 
the following $SU(2)_{c}$ violating terms:
\begin{eqnarray}\label{equ: coeffs from gauge boson sector}
\lefteqn{\frac{3}{16f^{2}}\cos{^{2}2\theta'}\left[\left(h_{1}^{\dag}D_{\mu}h_{1}\right)^2+\left(h_{2}^{\dag}D_{\mu}h_{2}\right)^2+2\left(h_{1}^{\dag}D_{\mu}h_{1}\right)\left(h_{2}^{\dag}D^{\mu}h_{2}\right)\right]}\nonumber\\
&+&\frac{1}{8f^{2}}\cos{^{2}2\theta}\left[\left(h_{1}^{\dag}D_{\mu}h_{2}\right)\left(h_{2}^{\dag}D^{\mu}h_{1}\right)-\left(h_{1}^{\dag}D_{\mu}h_{1}\right)\left(h_{2}^{\dag}D^{\mu}h_{2}\right)\right]+\mathrm{h.c.}
\end{eqnarray}
It seems surprising that there is a contribution from the heavy
$W$ at all ($\cos^2{2\theta}$ term) since its coupling is custodial
$SU(2)$-symmetric! The relevant operators appear with a relative 
minus sign, however, and cancel when we break
electroweak symmetry, giving a total contribution to the $T$
parameter of $1.6\left(\frac{1\TeV}{f}\right)^2\cos^2{2\theta'}$.  This 
mechanism is responsible for some more fortuitous cancellation in 
the next section.

At first glance it seems like we can minimize the $B'$
contribution to precision measurements by varying $\theta'$.
However we are constrained to $\sin{\theta'}\lesssim1/3$ by the relation
\begin{equation*}
\tan{\theta_W}=\frac{1}{q}\frac{\sin{\theta'}}{\sin{\theta}}
\end{equation*}
which gives us an unacceptably high $T$ parameter as well as 
large corrections to $G_F$ from $B'$ exchange.  To overcome
this problem the MM can be modified by replacing the
$SU(3)\times SU(2)\times U(1)$ gauge symmetry by $[SU(2)\times
U(1)]^2$ whose generators can be embedded into $SU(3)$ as
$\mathbf{T}^{1,2,3}$ and $\mathbf{T}^{8}$. This sidesteps the constraint, since we now have enough freedom to vary
$\theta'$ independently of $\theta$. If we charge the fermions
under $G_{L}$ as before, we will still have to tolerate large
non-oblique corrections. Altering the fermion couplings, 
however, by charging them under both U(1)s, gives a $B'$-
fermion coupling of:
\begin{equation}
ig'\sum_i{\overline{f}_i
\left(\frac{q_L}{\tan{\theta'}}-q_R\tan{\theta'}\right)\overline{\sigma}^{\mu}B'_{\mu}f_i}
\end{equation}
where $q_{L}$ and $q_{R}$ are the fermion charges under each of
the U(1)s.  We can set $q_{L}=q_{R}=q_{SM}/2$ for the
light fermions to eliminate this coupling at $\theta'\simeq\pi/4$,
provided we adjust the light yukawa couplings to account for the
new gauge structure:
\begin{eqnarray}\label{equ: Modified Yukawa couplings}
\mathcal{L}_{up}&=&\left[\lambda_{U}\left(\begin{array}{ccc}0&0&u^c\end{array}\right)X_{1}X_{4}^{\dag}\left(\begin{array}{c}q\\0\end{array}\right)\right][X_{33}]^{-\frac{3}{4}}+\hc\\
\mathcal{L}_{down}&=&\left[\lambda_{D}\left(\begin{array}{cc}q&0\end{array}\right)X_{1}X_{4}^{\dag}\left(\begin{array}{c}0\\0\\d^{c}\end{array}\right)+\lambda_{E}\left(\begin{array}{cc}l&0\end{array}\right)X_{1}X_{4}^{\dag}\left(\begin{array}{c}0\\0\\e^{c}\end{array}\right)\right][X_{33}]^{\frac{3}{4}}+\hc\nonumber
\end{eqnarray}
where $X_{33}$ is the 33 component of any of the link fields.  

Gauging an $SU(2)\times U(1)$ at both sites gives rise to an extra 
Higgs doublet, $h_w$, which is no longer eaten by gauge
bosons.  Its mass is zero at tree level, but its one-loop effective 
potential contains a logarithmically divergent contribution that
is of the same order as that of $h_1$ and $h_2$.  Since it is not 
coupled to the fermion sector or the Little Higgses, though, it does 
not pick up a vev.  We can therefore avoid the complications of 
working with three Higgs doublets in favor of just two.

\section{Non-linear Sigma Model Sector}

$SU(2)_{c}$ violating operators are also contained in the link field 
kinetic terms. It is straightforward to show that these are generated 
with the following coefficients:
\begin{eqnarray}\label{equ: nlsm ops}
\lefteqn{\frac{1}{16f^{2}}\left[\left(h_{1}^{\dag}D_{\mu}h_{1}\right)^2+\left(h_{2}^{\dag}D_{\mu}h_{2}\right)^2+2\left(h_{1}^{\dag}D_{\mu}h_{1}\right)\left(h_{2}^{\dag}D^{\mu}h_{2}\right)\right]}\nonumber\\
&+&\frac{1}{16f^{2}}\left[2\left(h_{1}^{\dag}D_{\mu}h_{2}\right)\left(h_{2}^{\dag}D^{\mu}h_{1}\right)-\left(h_{1}^{\dag}D_{\mu}h_{2}\right)^2-\left(h_{2}^{\dag}D_{\mu}h_{1}\right)^2\right]+\mathrm{h.c.}
\end{eqnarray}
Like the operators that originate from integrating out $W'$, the
terms in the second bracket will not give any contribution to 
the $T$ parameter after EWSB.  The contribution from the 
first bracket is $0.53\left(\frac{1\TeV}{f}\right)^2$.

\section{Plaquette Terms}
\label{Sec: plaquettes}

For an analysis of  the plaquette terms we use the 
Baker-Campbell-Hausdorff prescription to expand them to quartic 
order in the light Higgs fields. The $Z_{4}$
symmetry of the theory simplifies things greatly: it gets
rid of the $z$ tadpole, for example, leaving a $z$ mass:
\begin{equation}\label{equ: z mass}
M^{2}_{z}=4f^{2}\Re(\kappa)+O(\epsilon)
\end{equation}
a tree level mass for the Higgses which stabilizes the flat
direction in the potential and triggers electroweak symmetry
breaking;
\begin{equation}\label{equ: tree level Higgs masses}
\frac{\sqrt{3}f^{2}}{4}\Im(\epsilon)\left(h_{1}^{\dag}h_{1}-h_{2}^{\dag}h_{2}\right)
\end{equation}
a $z$-Higgs coupling of the form, $j^{a}z^{a}$, with 
\begin{equation}\label{equ: z current}
j^{a}=-\frac{f}{2}\Im(\epsilon)\Tr\left(\mathbf{T}^{a}[x,[x,\mathbf{T}^{8}]]-\mathbf{T}^{a}[y,[y,\mathbf{T}^{8}]]\right)+...
\end{equation}
and the leading quartic Higgs interaction
\begin{equation}\label{equ: quartic Higgs interaction}
\Re(\kappa)\Tr\left[x,y\right]^{2}
\end{equation}
We will neglect the T contribution from integrating out 
the heavy z since this is $O(\epsilon^2/\kappa^2)$ and so is 
suppressed by a factor of 100 in relation to the other terms
considered.

\section{Electroweak Symmetry Breaking}
\label{Sec: EWSB}
The leading order terms in the Higgs
potential (in manifestly CP invariant form) are
\begin{eqnarray}\label{equ: Higgs potential}
\lefteqn{V\approx
m_{1}^{2}h_{1}^{\dag}h_{1}+m_{2}^{2}h_{2}^{\dag}h_{2}+m_{12}^{2}\left(h_{1}^{\dag}h_{2}+h_{2}^{\dag}h_{1}\right)}\\
&+&\lambda_h\bigg[\left(h_{1}^{\dag}h_{1}\right)^{2}+\left(h_{2}^{\dag}h_{2}\right)^{2}-\left(h_{1}^{\dag}h_{1}\right)\left(h_{2}^{\dag}h_{2}\right)-\left(h_{1}^{\dag}h_{2}\right)\left(h_{2}^{\dag}h_{1}\right)\bigg]\nonumber
\end{eqnarray}
where the couplings include radiative corrections
as well as the tree level terms detailed in the
previous section. We are unable to say anything more precise since
two loop radiative corrections to the Higgs mass terms, for example, are
parametrically of the same order as one loop corrections.  We can,
however, place some constraints on the relative values of these
by imposing that the potential go to positive infinity far from
the origin. The quartic terms will dominate in this limit, but there is a
flat direction, namely $h_{1}=e^{i\varphi}h_{2}$ in which we demand
that the quadratic part of the potential be positive definite.  This gives 
us the constraint
\begin{equation}\label{equ: first constraint}
m_{1}^{2}+m_{2}^{2}\ge 2|m_{12}^2|
\end{equation}
Further requiring that the mass matrix for $h_{1}$ and $h_{2}$ have
one negative eigenvalue at the origin tells us that
\begin{equation} \label{equ: second constraint}
m_{1}^{2}m_{2}^{2}<m_{12}^{4}
\end{equation}
The potential (Equation \ref{equ: Higgs potential}) is minimized for vevs 
of the form
\begin{eqnarray}\label{equ: vevs}
h_{1}=\frac{1}{\sqrt{2}}\left(\begin{array}{c}0\\v\cos{\beta}\end{array}\right)\nonumber\\
h_{2}=\frac{1}{\sqrt{2}}\left(\begin{array}{c}0\\v\sin{\beta}\end{array}\right)
\end{eqnarray}
where
\begin{eqnarray}\label{equ: minimum}
v^{2}&=&\frac{1}{\lambda_h}\left[-m_{1}^{2}-m_{2}^{2}+\frac{|m_{1}^{2}-m_{2}^{2}|}{\cos{2\beta}}\right]\nonumber\\
\sin{2\beta}&=&-\frac{2m_{12}^{2}}{m_{1}^{2}+m_{2}^{2}}
\end{eqnarray}
An examination of the solution shows that it is consistent with
the constraints (\ref{equ: first constraint}) and (\ref{equ:
second constraint}).

The masses of the physical states in the two-doublet sector satisfy the relations
\begin{eqnarray} \label{equ: relation between masses}
4m^{2}_{H^{\pm}}&=&m^{2}_{h^{0}}+m^{2}_{H^{0}}+3m^{2}_{A^{0}}\\
m^{2}_{H^{\pm}}&=&m^{2}_{A^{0}}+\lambda_{h} v^{2}\nonumber
\end{eqnarray}

\section{Fermion Sector}
\label{Sec: fermion loops}
Armed with this information we can now calculate the
$T$ and $S$ parameters from the fermion sector.  We look directly at
corrections to the $W$ and $Z$ masses from vacuum polarization
diagrams containing fermion loops.  

The Higgses give rise to a small mixing term for
the top and heavy fermion in our theory so we need to find the
fermion mass eigenstates. Diagonalizing the Yukawa coupling in two
stages: to zeroth order in $v$ to start with, we get $\mathcal{L}_t$ in
terms of the new eigenstates:
\begin{eqnarray}\label{equ: fermion rotation1}
\mathcal{L}_{t}=f\sqrt{\lambda^{2}+\lambda^{'2}}\bigg[\widetilde{U}^{c}U+\sin^{2}{\xi}\left(\begin{array}{ccc}0 & 0 & \widetilde{U}^{c}\end{array}\right)\left(X_{1}X_{4}^{\dag}-1\right)\left(\begin{array}{c}q_{3}\\U\end{array}\right)\\
+\sin{\xi} \cos{\xi} \left(\begin{array}{ccc} 0 & 0 & u^{c}_{3}\end{array}\right)\left(X_{1}X_{4}^{\dag}-1\right)\left(\begin{array}{c}q_{3}\\U\end{array}\right)\bigg]\nonumber
\end{eqnarray}
where
\begin{eqnarray}\label{equ: xi}
\sin{\xi}&=&\frac{\lambda}{\sqrt{\lambda^{2}+\lambda^{'2}}}\nonumber\\
\widetilde{U}^{c}&=&\cos{\xi} U^{c}+\sin{\xi} u_{3}^{c'}\\
u_{3}^{c}&=&-\sin{\xi} U^{c}+\cos{\xi} u_{3}^{c'}\nonumber
\end{eqnarray}
Expanding the link fields to first order in v/f, a convenient phase rotation 
gives us the following terms in the $t-U$ mass 
matrix:
\begin{eqnarray}\label{fermionmassmatrix}
m_{tt}&=&\sqrt{\lambda^{2}+\lambda'^{2}}\sin{\xi}\cos{\xi}\frac{v}{\sqrt{2}}(\sin{\beta}+\cos{\beta})\nonumber\\
m_{tU}&=&\sqrt{\lambda^{2}+\lambda'^{2}}\sin^{2}{\xi}\frac{v}{\sqrt{2}}(\sin{\beta}+\cos{\beta})\\
m_{UU}&=&f\sqrt{\lambda^2+\lambda'^2}\nonumber
\end{eqnarray}
Using $m_{tt},m_{tU}<<m_{UU}$,
$m_{Ut}=0$ we approximate the results in Appendix \ref{Sec:
appendix} to obtain
\begin{eqnarray}\label{equ: fermion mass eigenstates and rotation
angle}
    m_{t}^{2}&\approx& m_{tt}^{2}\nonumber\\
    m_{U}^{2}&\approx&
    m_{UU}^{2}\left(1+\frac{m_{tU}^{2}}{m_{UU}^{2}}\right)\\
    \cos{\theta_{L}}&\approx& 1-\frac{m_{tU}^{2}}{2m_{UU}^{2}}\nonumber
\end{eqnarray}
Now we can fix the top Yukawa coupling to its value $\lambda_{t}$ in the SM, 
which for a given value of $\tan{\beta}$ relates $\lambda$ to $\lambda'$ in the
following way:
\begin{equation} \label{equ: relation between lambdas}
\frac{\lambda'}{\lambda}=\frac{\lambda_{t}\sqrt{1+\tan^2{\beta}}}{\sqrt{\lambda^{2}(1+\tan{\beta})^{2}-\lambda_{t}^{2}(1+\tan^{2}\beta)}}
\end{equation}
with $\lambda$ constrained by
\begin{equation} \label{equ: lambda constraint}
\lambda^{2}>\frac{1+\tan^{2}\beta}{(1+\tan{\beta})^{2}}\lambda_t^2
\end{equation}
Having determined the fermion mass eigenvalues, we use \cite{Lavoura:1993np} 
to find:
\begin{eqnarray}\label{T and S from fermion sector}
T_{f}=\frac{3}{16\pi\sin^2{\theta_{W}}\cos^2{\theta_{W}}}\bigg[\sin^2{\theta_{L}}\Theta_{+}\left(\frac{m_{U}^{2}}{m_{Z}^{2}},\frac{m_{b}^2}{m_{Z}^{2}}\right)-\sin^2{\theta_{L}}\Theta_{+}\left(\frac{m_{t}^{2}}{m_{Z}^{2}},\frac{m_{b}^{2}}{m_{Z}^{2}}\right)\hskip3cm\\
\hskip6.7cm-\sin^2{\theta_{L}}\cos^2{\theta_{L}}\Theta_{+}\left(\frac{m_{U}^{2}}{m_{Z}^{2}},\frac{m_{t}^2}{m_{Z}^{2}}\right)\bigg]\nonumber\\
S_{f}=\frac{3}{2\pi}\left[\sin^2{\theta_{L}}\Psi_{+}\left(\frac{m_{U}^{2}}{m_{Z}^{2}},\frac{m_{b}^2}{m_{Z}^{2}}\right)-\sin^2{\theta_{L}}\Psi_{+}\left(\frac{m_{t}^{2}}{m_{Z}^{2}},\frac{m_{b}^2}{m_{Z}^{2}}\right)-\sin^2{\theta_{L}}\cos^2{\theta_{L}}\chi_{+}\left(\frac{m_{U}^{2}}{m_{Z}^{2}},\frac{m_{t}^2}{m_{Z}^{2}}\right)\right]\nonumber
\end{eqnarray}
for
\begin{eqnarray}\label{function definitions}
\Theta_{+}(y_1,y_2)&=&y_1+y_2-\frac{2y_1y_2}{y_1-y_2}\ln{\frac{y_1}{y_2}}\quad\quad \Psi_{+}(y_1,y_2)=\frac{1}{3}-\frac{1}{9}\ln{\frac{y_1}{y_2}}\nonumber\\
\chi_{+}(y_1,y_2)&=&\frac{5(y_1^2+y_2^2)-22 y_1 y_2}{9(y_1-y_2)^2}+\frac{3 y_1 y_2(y_1+y_2)- y_1^3- y_2^3}{3(y_1-y_2)^3}\ln{\frac{y_1}{y_2}}
\end{eqnarray}

\section{Higgs Sector}
\label{Sec: higgs loops}
There is a contribution to the vacuum polarization diagrams 
from additional physical Higgs states running around the loop.  
This is a standard calculation (see \cite{Gregoire:2003kr,
Gunion:1989we}) which yields
\begin{eqnarray}
\nonumber
T_h=\frac{1}{16\pi\sin^2{\theta_W}m^2_{W}}\Big( F(m^2_{A^0},m^2_{H^\pm}) + \cos^2(\alpha -\beta) \big(F(m^2_{H^\pm}, m^2_{h^0})- F(m^2_{A^0}, m^2_{h^0}) \big)\\
\hskip4cm+ \sin^2(\alpha -\beta) \big( F(m^2_{H^\pm}, m^2_{H^0})-F(m^2_{A^0}, m^2_{H^0})\big) \Big)\nonumber\\
S_h= \frac{1}{12 \pi}\Big( \cos^2(\beta -\alpha)\log\frac{m^2_{H^0}}{m^2_{h^0}}- \frac{11}{6} + \sin^2(\beta -\alpha) G(m^2_{H^0}, m^2_{A^0},m^2_{H^\pm})\hskip2cm\\
\hskip6cm +\cos^2(\beta -\alpha) G(m^2_{h^0}, m^2_{A^0},m^2_{H^\pm})\Big)\nonumber
\end{eqnarray}
where
\begin{eqnarray}
F(x,y) &=& \frac{1}{2}(x + y) - \frac{xy}{x -y} \log\frac{x}{y}\nonumber\\
G(x,y,z)& =& \frac{x^2 + y^2}{(x - y)^2} + \frac{(x- 3 y)x^2\log\frac{x}{z} - (y - 3x) y^2 \log \frac{y}{z}}{(x-y)^3}\nonumber
\end{eqnarray}
The $A$, $H$, $h$ are Higgs mass eigenstates and $\alpha$ is the
mixing angle between $H^0$ and $h^0$, as detailed in
\cite{Gunion:1989we}.

\section{Results}
\label{Sec: results}

The graphs below give some idea of the size of oblique corrections from the 
Higgs and fermion sectors.  It can be seen in Figure \ref{Fig: fermions} that 
the $T$ parameter contribution from fermions is rather small ($S$ is negligible) for the most part, and 
decreases with increasing $\tan{\beta}$.  However the top partner also 
gets heavier in this limit, increasing the level of fine tuning in the theory, since 
the quadratically divergent fermion loop diagram is cut off at a higher energy.
\FIGURE[h]{\epsfig{file=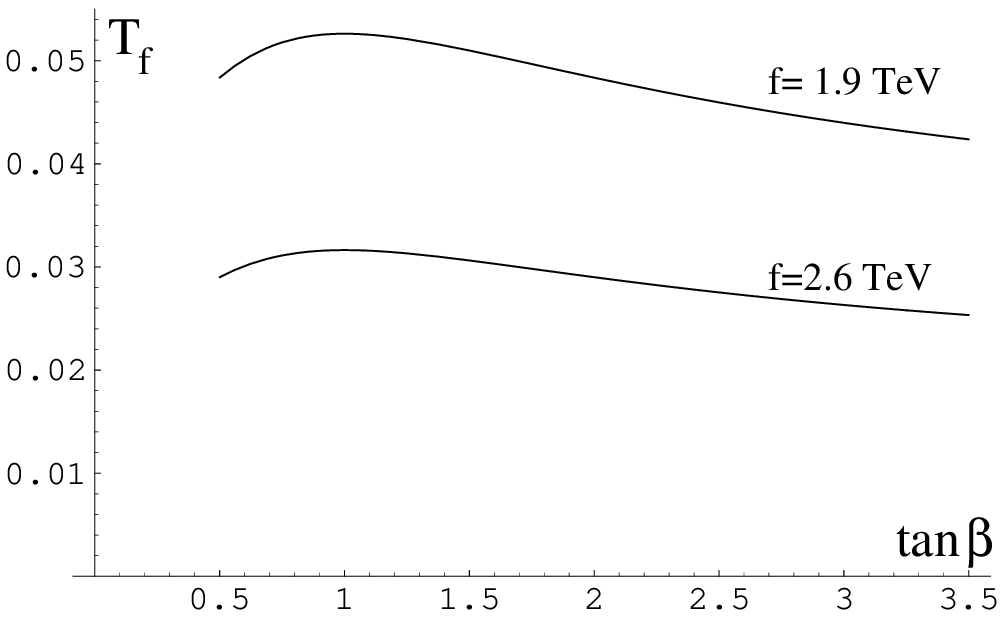, width =
8cm}\epsfig{figure=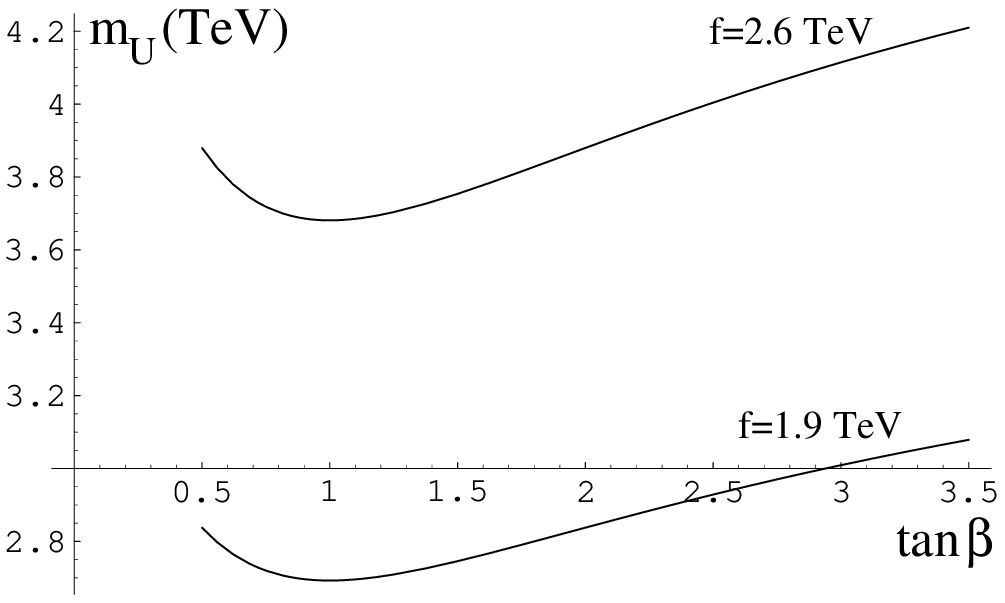,width=8cm}\caption{Fermion 
sector contribution to $T$ and mass of top partner as a function of $\tan{\beta}$.
$\lambda$ and $\lambda'$ were chosen to minimize $m_U$ with a fixed top 
quark mass.}\label{Fig: fermions}}

The Higgs sector contribution to the $T$ parameter is generically negative, 
although there is no such restriction on the $S$ parameter (see Figure 
\ref{Fig: Higgs sector}).  As for the fermions, though, the 
latter is usually small and can be ignored .
\FIGURE[h]{\epsfig{figure=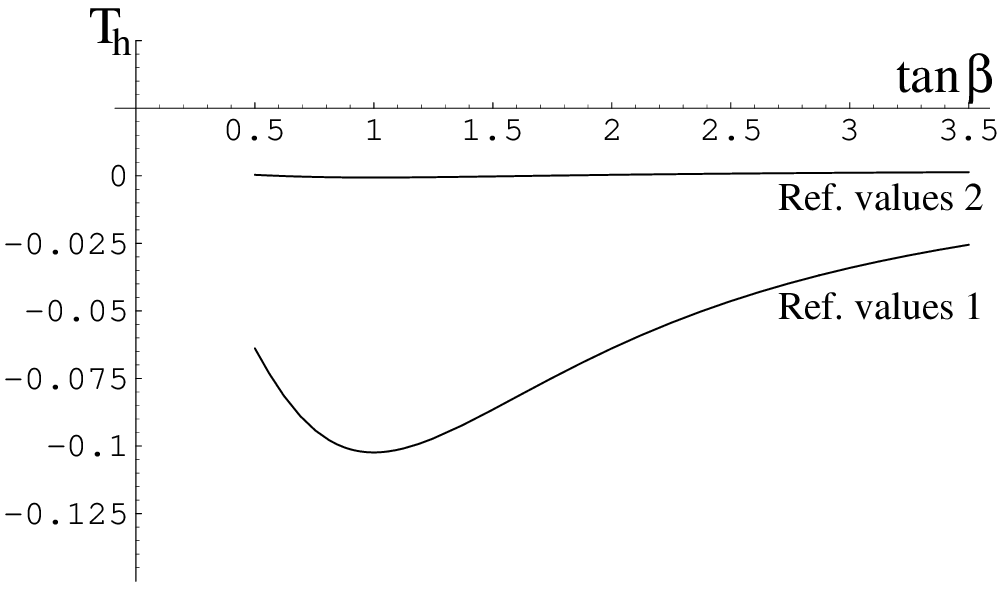, width = 7.2cm}
\epsfig{figure=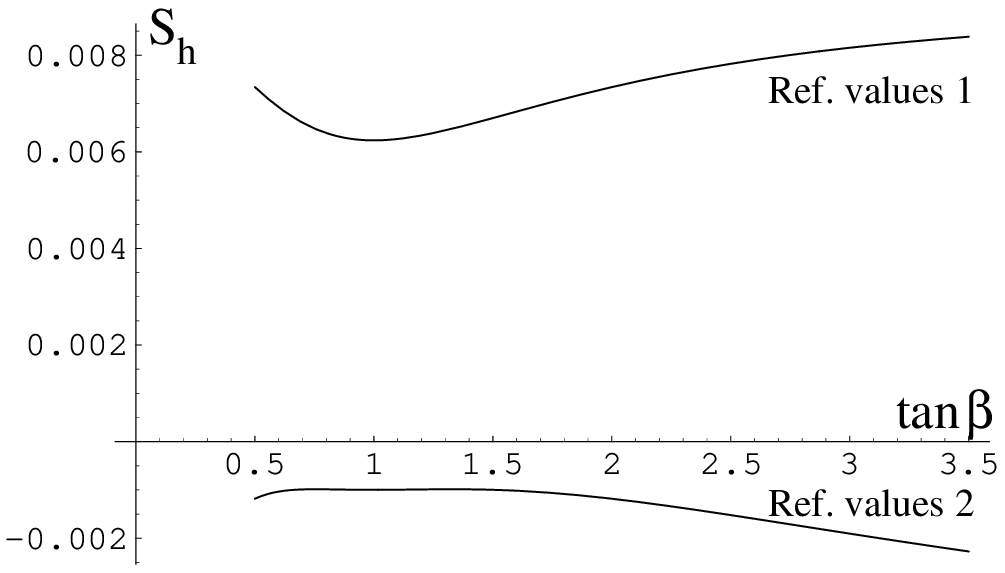, width=7.2cm} \caption{Higgs sector contribution to PEWOs as a function 
of $\tan{\beta}$.  Values of other variables taken from Table \ref{Tab: parameters}} \label{Fig: Higgs sector}}
The biggest constraint in our models is the large
$T$ parameter arising in the nonlinear sigma model sector.  Keeping 
this at a manageable level limits us to $f\gtrsim 2$ TeV.  At this breaking scale the
remaining parameters can have a range of values
that do not take us beyond 1.5-$\sigma$ in the $S$-$T$
plane.  To illustrate this we chose two representative sets of free
parameters (Table \ref{Tab: parameters}) and plotted $T$ against $S$, subtracting out  the SM $T$ and $S$ contributions.  The first reference set (see Figure \ref{Fig: parameters1}), which contains a moderately heavy Higgs, has parameters which were chosen to obtain a sizable negative $T$ from the Higgs sector to partly cancels the nonlinear sigma model contribution, thus allowing us to make $f$ as low as $1.9$ TeV without leaving the ellipse.  We also plot the fine tuning for different regions of parameter space within the ellipse in Figure \ref{Fig:r1v} by varying the Higgs quartic coupling and $\tan{\beta}$ around this reference set.  One can see that there are viable regions with larger quartic coupling which can give even less fine tuning in the Higgs mass, however these lie in a smaller band in parameter space and thus correspond to a more specific choice of the parameters of the theory. In fact, the allowed region ends for large values of the quartic coupling because it is driven out of the ellipse by a $T$ contribution from the Higgs sector that is too negative. One could imagine taking an even smaller value for $f$ and thus increasing the positive nonlinear sigma model contribution to $T$, expanding the allowed region and decreasing the fine tuning further, since the Higgs mass is increased as the top quark partner mass is decreased.  However, as before, this occurs for more and more specific choices of parameters where large $T$ contributions from the nonlinear sigma model and Higgs sectors are delicately cancelling out  and we chose not to work with such values.
Our second set of parameters (see Figure \ref{Fig: parameters2}), which was picked to contain a light Higgs but is otherwise fairly random, takes us only slightly out of the $S$-$T$ ellipse.  It has a small negative $S$ and no cancellation between 
sectors, which forces us to choose a larger value for $f$.  We vary $\lambda_{h}$ and $\tan{\beta}$  around this reference set in Figure \ref{Fig:r2v}.  We see that there is a large region of parameter space where the PEWOs are no further outside the 1.5$\sigma$ ellipse than the reference point we chose, and the fine tuning is even better.  Since $S$ and $T$ are not as sensitive to the other parameters one can conclude that the results we quote are quite generic in the parameter space of the model.  Note that although the theory seems to favour a heavy Higgs, it is still possible to 
find acceptable data sets in which it is light. 

More generically consistency with PEWOs constrains us to values of $f$ greater than 2 TeV.  The increase of the heavy quark mass with  $f$ bounds the latter to be less than 2.5 TeV for the fine tuning to be any better than that of the SM. The acceptable region in parameter space is larger for higher values of $f$, however this comes with the price of increased fine tuning in the higgs mass.
\TABLE[h]{
\begin{tabular}{|c|c|c|}\hline
{Parameter}&{Reference values 1} & {Reference values 2} \\
  \hline \hline
 $f$(TeV) & $1.9$ & $2.6$\\
\hline $\theta$ & $40^\circ$ & $25^\circ$\\
\hline $\theta'$ & $47^\circ$ & $50^\circ$\\
\hline $\lambda$ & $0.9$ & $1.1$\\
\hline $\lambda_h$ & $1.6$ & $0.5$\\
\hline $\tan{\beta}$ & $1.1$ & $2.0$\\
\hline $m_{H^\pm}$(GeV) & $234$ & $206$ \\
\hline $m_{H^0}$(GeV) & $381$ & $206$\\
\hline $m_{A^0}$(GeV) & $98$ & $191$\\
\hline $m_{h_0}$(GeV) & $220$ & $134$\\
\hline $m_U$(TeV) & $2.77$ & $3.89$\\
\hline $m_W'$(TeV) & $2.56$ & $4.50$\\
\hline $m_B'$(TeV) & $0.78$ & $1.08$\\
\hline fine tuning & $17\%$ & $3\%$\\
\hline
\end{tabular}
\caption{\label{Tab: parameters} Two sets of reference parameters
for the Modified Minimal Moose.}}
\FIGURE[h]{ \epsfig{file=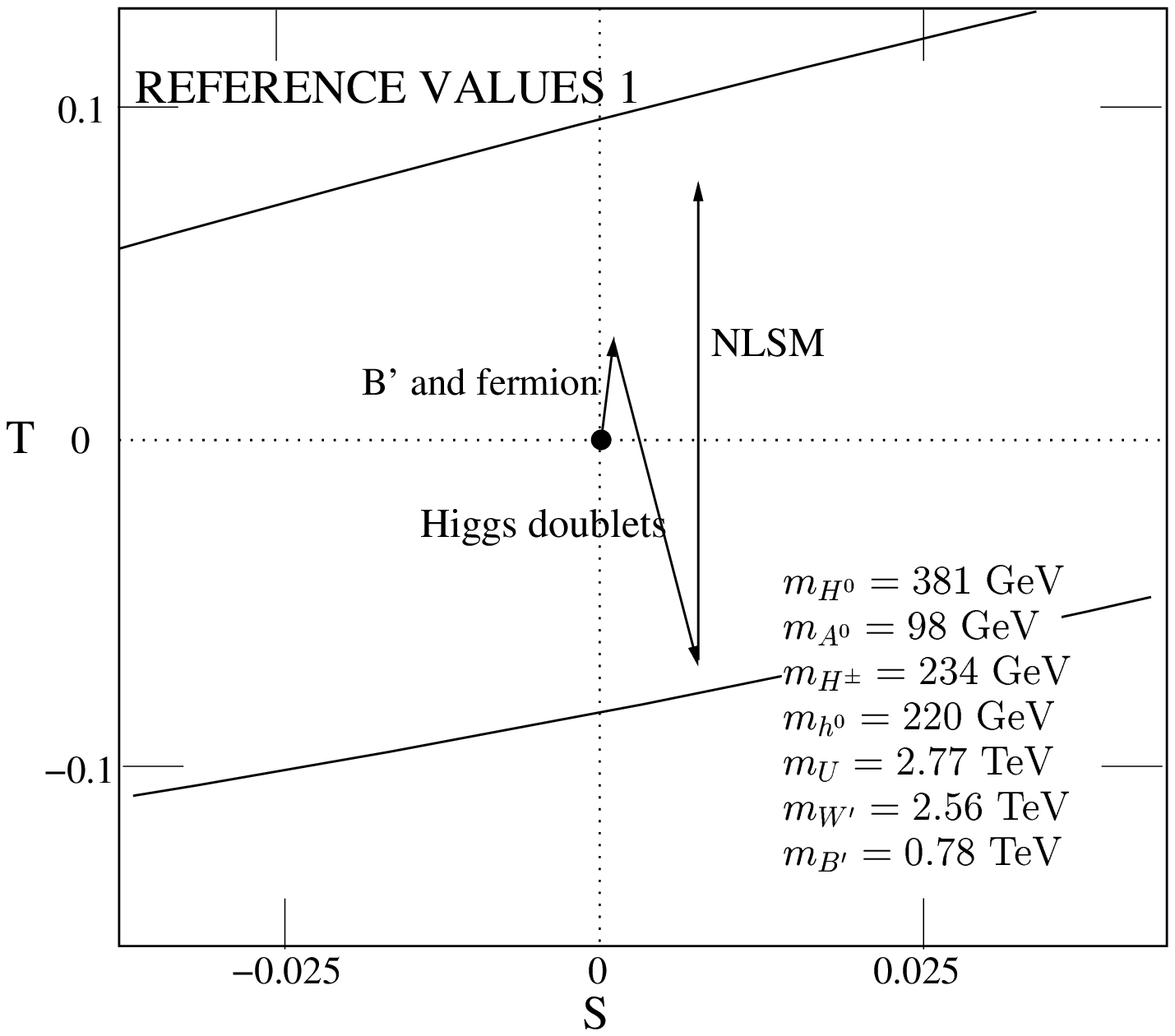}\caption{$S$ and $T$ values in the MMM for reference set 1 in Table \ref{Tab:
parameters} plotted on a 1.5-$\sigma$ oval in the S-T plane}\label{Fig: parameters1}}
\FIGURE[h]{ \epsfig{file=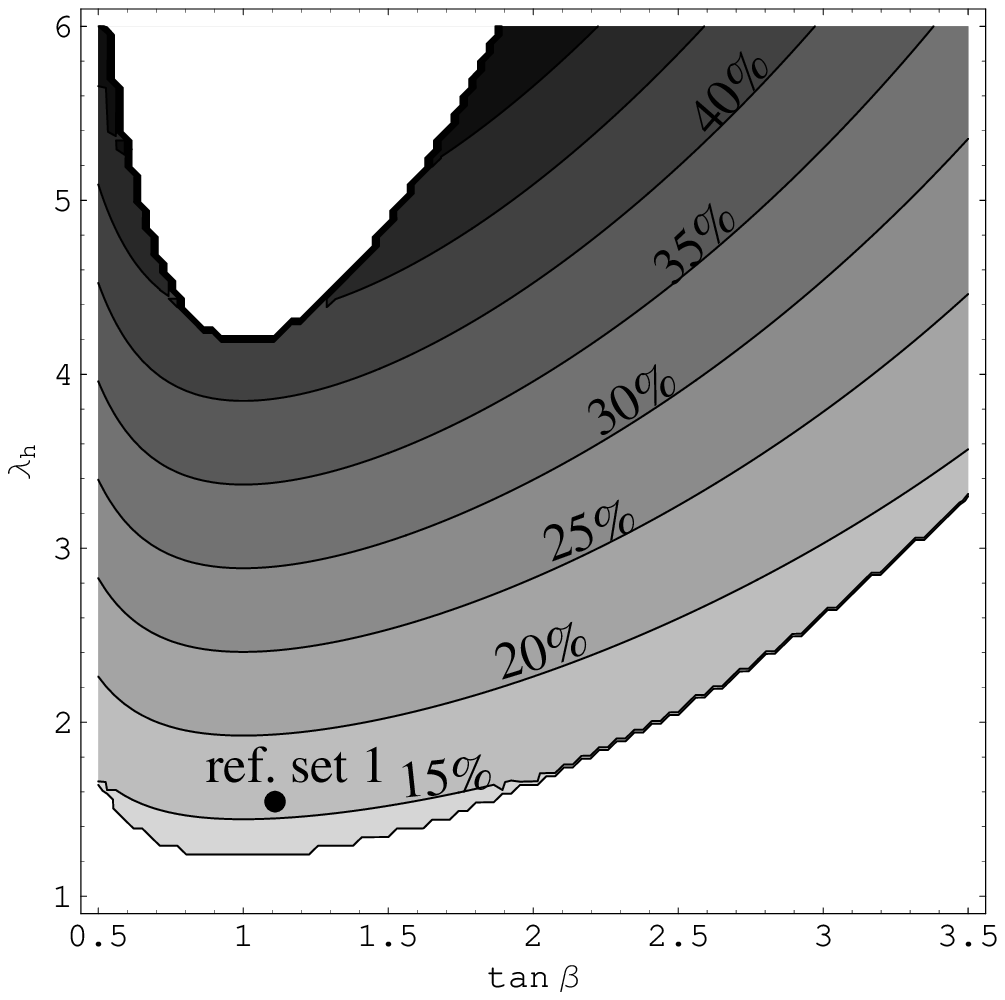}\caption{Fine tuning plotted as a function of $\tan(\beta)$ and $\lambda_{h}$ where all other parameters have been taken from Table \ref{Tab: parameters}. We only plot regions of parameter space which lie within the $S$-$T$ ellipse. We indicate the position of reference set 1. Note that the fine tuning improves for larger values of $\lambda_{h}$ as the Higgs becomes heavier.} \label{Fig:r1v}}
\FIGURE[h]{ \epsfig{file=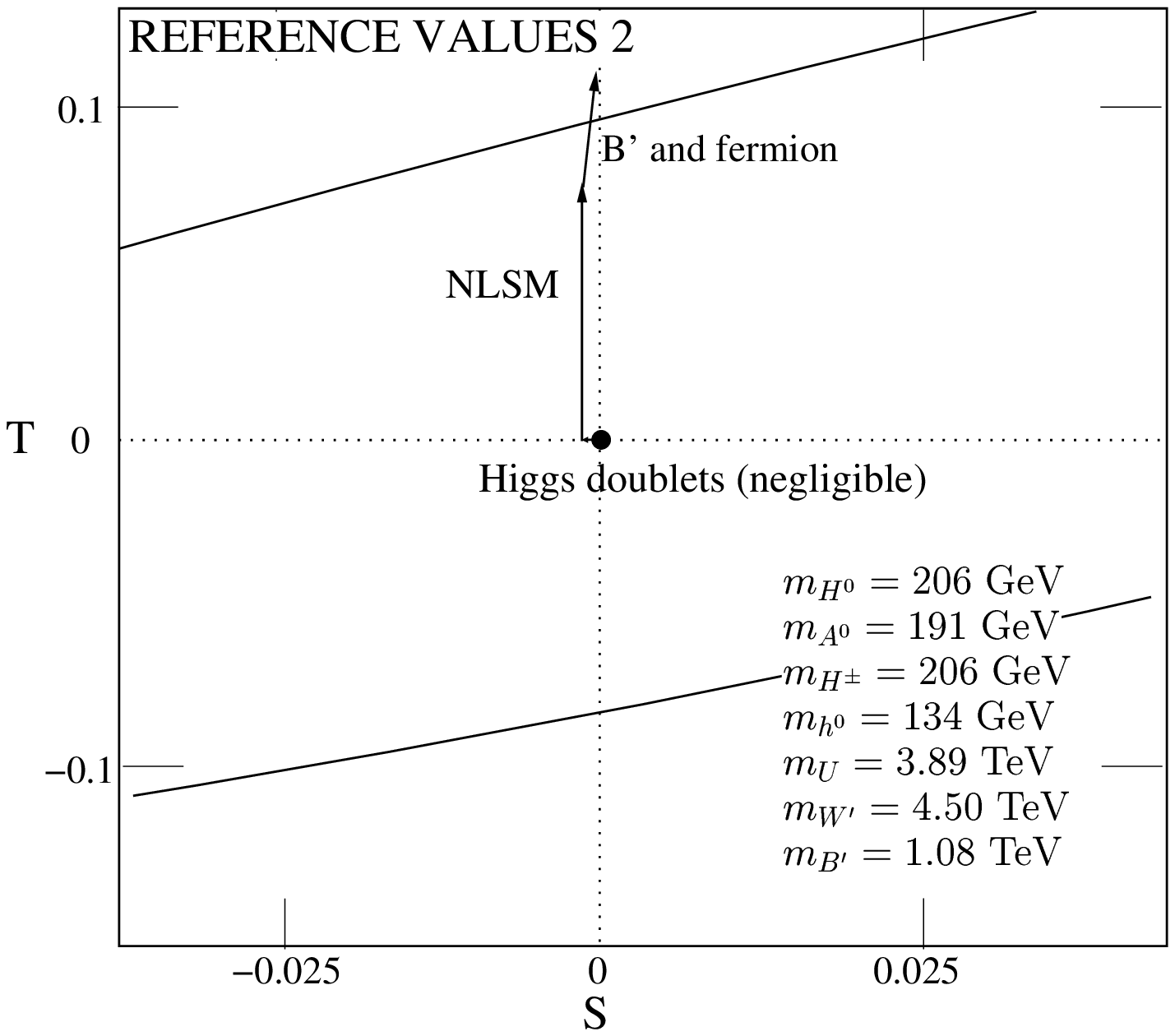}\caption{$S$ and $T$ values in the MMM for reference set 2 in Table \ref{Tab:
parameters} plotted on a 1.5-$\sigma$ oval in the S-T plane}\label{Fig: parameters2}}
\FIGURE[h]{ \epsfig{file=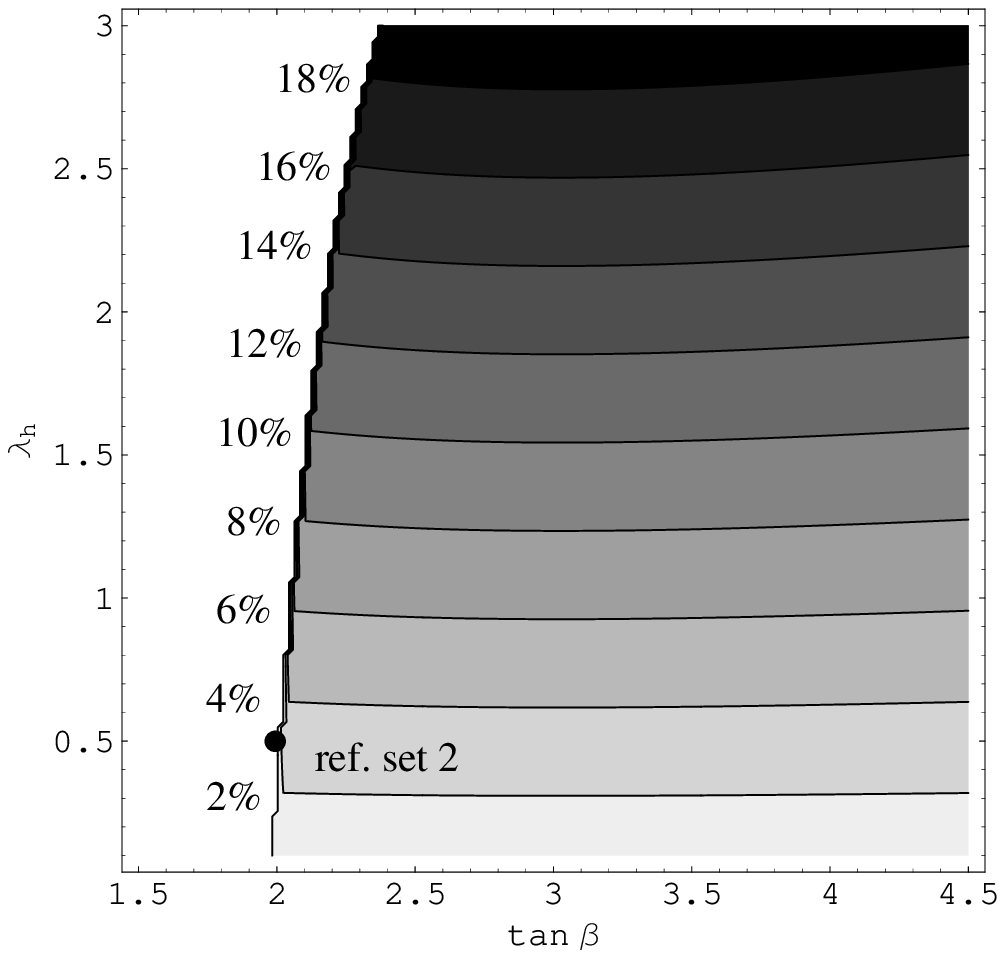}\caption{Fine tuning plotted as a function of $\tan{\beta}$ and $\lambda_{h}$ where all other parameters have been taken from Table \ref{Tab: parameters}. We plot regions of parameter space which are at least as close to the $S$-$T$ ellipse as reference set 2.} \label{Fig:r2v}}
         
\section{Conclusion}

Little Higgs models, like the Minimal Moose, predict new heavy
particles at the TeV scale.  Upon integrating out these particles 
the SM is recovered at low energies, with possibly one or more 
extra Higgs doublets.  At higher energies the Higgses form 
components of nonlinear sigma model fields which become 
strongly coupled at around 10 TeV.  At still higher energies a 
UV completion of the theory is needed. This could be achieved 
with strongly coupled dynamics \cite{Nelson:2003aj}, a linear sigma
model or supersymmetry. In the latter case the SUSY breaking scale 
is pushed to 10 TeV, alleviating the difficulties of flavor-changing 
neutral currents associated with TeV-scale superpartners.

At energies below the masses of these new particles we rely on
precision electroweak data to gauge the feasibility of a
particular model as a possible extension to the SM.  In the
absence of new flavor physics (due to the introduction of
a partner for the top quark only), precision measurements
can be divided into oblique and non-oblique corrections.  We
analyzed these for two such models at around a TeV, translating
the low energy theory into effective operator language as far as
possible. We saw that we ran into significant problems in more
than one sector when we considered the constrained gauge structure
of the MM.  Gauging two copies of $SU(2)\times U(1)$ instead and
charging the SM fermions equally under both U(1)s, as in the MMM,
does away with these issues as we can then go to the near oblique
limit without reintroducing large contributions to the $T$ parameter
from $B'$ exchange.  

This might be understood better in the context of other LH theories, the 
Littlest Higgs \cite{Arkani-Hamed:2002qy}
for example.  The greatest contrast between this and the MM is the nonlinear sigma model
sector, where the Littlest Higgs has a built-in $SU(2)_{c}$ symmetry
which protects it from any $T$ parameter contribution.  This symmetry is explicitly 
broken in the top sector, but only by a small amount.  The gauge sectors of the 
theories are identical except for a $B'$ mass in the Littlest Higgs 
which is lighter by a factor of 2 (since the theory only contains 1 link field), but heavier
by $\sqrt{5/3}$ to account for the different group structure involved.  
Aside from this, the similarity in the general framework of the models
 implies that a lower cutoff can be tolerated in the case of the Littlest Higgs, 
giving rise to lower masses for the heavy particles, and a subsequent decrease
in the level of fine tuning.  The relative success of the MM is rather surprising, 
however, given that it was designed for minimality rather than freedom from 
precision electroweak constraints.

In summary we see that the MMM, which contains a gauged $\left[SU(2)\times U(1)\right]^2$ is a viable candidate for TeV-scale
physics.  The heavy counterparts for SM particles give
rise to precision electroweak corrections that are within
acceptable experimental bounds for a large range of parameters of
the theory.  It leads to at least moderate improvements over the SM in terms of the gauge hierarchy problem for generic regions of parameter space, and very significant improvements for less generic regions, which are nevertheless plausibly large.  It remains for the LHC to confirm whether there is a role
for Little Higgs theories in physics beyond the Standard
Model.

\section*{Acknowledgments}
We would like to thank Nima Arkani-Hamed for his many effective field
theory indoctrination sessions among other things; Jay Wacker,
without whose huge reserves of patience none of this would have been 
possible; and Spencer Chang and Thomas Gregoire for many helpful hints.

\appendix
\section{Top Seesaw}
\label{Sec: appendix}

Starting with a top sector mass term of the form
\begin{equation}\label{equ: fermion mass matrix}
    -\left(\begin{array}{cc} \overline{t} & \overline{U}
    \end{array}\right) \left(\begin{array}{cc} m_{tt} & m_{tU}\\m_{Ut} &
    m_{UU}\end{array}\right)\left(\begin{array}{c} \overline{t^{c}} \\
    \overline{U^{c}}\end{array}\right)
\end{equation}
we can diagonalize the $M^{\dag}M$ matrix as in
\cite{Chivukula:1998wd} giving mass eigenvalues of
\begin{equation}\label{equ: fermion eigenvalues}
m_{t,U}^{2}=\frac{1}{2}\left[m_{UU}^{2}+m_{tt}^{2}+m_{Ut}^{2}+m_{tU}^{2}\pm\sqrt{(m_{UU}^{2}+m_{tt}^{2}+m_{Ut}^{2}+m_{tU}^{2})^{2}-4(m_{UU}m_{tt}-m_{tU}m_{Ut})^{2}}\right]
\end{equation}
with different mixing angles on the right and left
\begin{eqnarray}\label{equ: fermion mixing angles}
    \left(\begin{array}{c}t'\\U'\end{array}\right)=\left(\begin{array}{cc}c_{L} & -s_{L}\\s_{L} & c_{L}\end{array}\right)\left(\begin{array}{c}t\\U\end{array}\right)\nonumber\\
    \left(\begin{array}{c}\overline{t_{c}}'\\\overline{U_{c}}'\end{array}\right)=\left(\begin{array}{cc}c_{R} & s_{R}\\-s_{R} & c_{R}\end{array}\right)\left(\begin{array}{c}-\overline{t_{c}}\\\overline{U_{c}}\end{array}\right)
\end{eqnarray}
and $c_{L}$ given by
\begin{equation}
\cos{\theta_{L}}=\frac{1}{\surd{2}}\left[1+\frac{m_{UU}^{2}-m_{tt}^{2}+m_{Ut}^{2}-m_{tU}^{2}}{m_{U}^{2}-m_{t}^{2}}\right]^{\frac{1}{2}}
\end{equation}

\nocite{*}
\bibliographystyle{JHEP}
\bibliography{PEWO}
\end{document}